\documentclass[preprint,aps,prc,tightenlines,nofootinbib,showpacs,showkeys]{revtex4}
\begin{document}
\input{psfig}
\input{epsf}
\def\Im{\mbox{\sl Im\ }}
\def\pd{\partial}
\def\oln{\overline}
\def\olft{\overleftarrow}
\def\ds{\displaystyle}
\def\bgreek#1{\mbox{\boldmath $#1$ \unboldmath}}
\def\sla#1{\slash \hspace{-2.5mm} #1}
\newcommand{\bra}{\langle}
\newcommand{\ket}{\rangle}
\newcommand{\vep}{\varepsilon}
\newcommand{\met}{{\mbox{\scriptsize met}}}
\newcommand{\lab}{{\mbox{\scriptsize lab}}}
\newcommand{\cm}{{\mbox{\scriptsize cm}}}
\newcommand{\mcal}{\mathcal}
\newcommand{\Del}{$\Delta$}
\newcommand{\g}{{\rm g}}
\long\def\Omit#1{}
\long\def\omit#1{\small #1}
\def\beq{\begin{equation}}
\def\eeq{\end{equation} }
\def\bea{\begin{eqnarray}}
\def\eea{\end{eqnarray}}
\def\eqref#1{Eq.~(\ref{eq:#1})}
\def\eqlab#1{\label{eq:#1}}
\def\figref#1{Fig.~\ref{fig:#1}}
\def\figlab#1{\label{fig:#1}}
\def\tabref#1{Table \ref{tab:#1}}
\def\tablab#1{\label{tab:#1}}
\def\secref#1{Section~\ref{sec:#1}}
\def\seclab#1{\label{sec:#1}}
\def\VYP#1#2#3{#1 (#2) #3}  
\def\NP#1#2#3{Nucl.~Phys.~\VYP{#1}{#2}{#3}}
\def\NPA#1#2#3{Nucl.~Phys.~A~\VYP{#1}{#2}{#3}}
\def\NPB#1#2#3{Nucl.~Phys.~B~\VYP{#1}{#2}{#3}}
\def\PL#1#2#3{Phys.~Lett.~\VYP{#1}{#2}{#3}}
\def\PLB#1#2#3{Phys.~Lett.~B~\VYP{#1}{#2}{#3}}
\def\PR#1#2#3{Phys.~Rev.~\VYP{#1}{#2}{#3}}
\def\PRC#1#2#3{Phys.~Rev.~C~\VYP{#1}{#2}{#3}}
\def\PRD#1#2#3{Phys.~Rev.~D~\VYP{#1}{#2}{#3}}
\def\PRL#1#2#3{Phys.~Rev.~Lett.~\VYP{#1}{#2}{#3}}
\def\FBS#1#2#3{Few-Body~Sys.~\VYP{#1}{#2}{#3}}
\def\AP#1#2#3{Ann.~of Phys.~\VYP{#1}{#2}{#3}}
\def\ZP#1#2#3{Z.\ Phys.\  \VYP{#1}{#2}{#3}}
\def\ZPA#1#2#3{Z.\ Phys.\ A\VYP{#1}{#2}{#3}}
\def\half{\mbox{\small{$\frac{1}{2}$}}}
\def\quarter{\mbox{\small{$\frac{1}{4}$}}}
\def\nn{\nonumber}
\newlength{\PicSize}
\newlength{\FormulaWidth}
\newlength{\DiagramWidth}
\newcommand{\vslash}[1]{#1 \hspace{-0.42 em} /}
\newcommand{\qslash}[1]{#1 \hspace{-0.46 em} /}
\def\her{\marginpar{$\Longleftarrow$}}
\def\bel{\marginpar{$\Downarrow$}}
\def\abo{\marginpar{$\Uparrow$}}



\title{Calculation of two-photon exchange effects for $\Delta$ production in
electron-proton collisions}

\author{S.~Kondratyuk}
\thanks{Corresponding author}
\email{kondrat@physics.umanitoba.ca}
\affiliation{Department of Physics and Astronomy, University of Manitoba,
Winnipeg, MB, Canada R3T 2N2}
\author{P.~G.~Blunden}
\affiliation{Department of Physics and Astronomy, University of Manitoba,
Winnipeg, MB, Canada R3T 2N2}

\date{\today}

\begin{abstract}

We present an evaluation of a two-photon exchange correction to the cross section for unpolarized
$\Delta$ isobar production in electron-proton collisions, using a relativistic,
crossing symmetric and gauge invariant approach. The calculated box and crossed-box diagrams include nucleon and $\Delta$ intermediate states. We find a relation between the angular nonlinearity of the two-photon exchange contribution and 
the value of the $\gamma \Delta \Delta$ coupling constant.

\end{abstract}

\maketitle


\section{Introduction} \seclab{intro}

The internal structure of hadrons has been extensively studied using
elastic and inelastic electron scattering. For a theoretical
understanding of such processes one traditionally relies on the
leading order (Born) contribution, an exchange of a single virtual photon
between the electron and the hadron. The generally suppressed higher-order terms, including two-photon exchange contributions,
have also been considered (see e.~g.~Refs.~\cite{Dre59,MoT69}). 
More recently, the particular importance of the two-photon exchange contributions has been emphasized~\cite{Blu03,Gui03,Che04,Kon05} since
these corrections are necessary to explain simultaneously two distinct types of 
electron-proton scattering experiments: the Rosenbluth separation and the 
polarization-transfer techniques. Therefore, our understanding of such fundamental quantities as the nucleon electromagnetic form factors cannot be complete without including two-photon exchange effects in the analyses of elastic electron-nucleon scattering.

Inelastic reactions allow one to obtain additional crucial information on hadronic structure.
Of particular interest is the situation where the final hadronic state
in an electron-proton collision lies in the $\Delta$ resonance region.    
This reaction has been recently studied both experimentally~\cite{Lia04,Tva05} and
theoretically~\cite{Pas06}.
From the theoretical point of view, it is important as one of the
simplest processes involving the electromagnetic nucleon-to-$\Delta$ transition current
($\gamma N \Delta$ vertex) which parametrizes the one-photon exchange amplitude.  
By analogy with the case of elastic scattering, it is obvious that
although the one-photon exchange mechanism is generally dominant, 
two-photon exchange effects in $\Delta$ production can be quite important 
and thus have to be carefully examined. 

This paper describes a calculation of 
two-photon exchange corrections to the unpolarized cross section for $\Delta$ production in electron-proton collisions.
We use a quantum field theoretical approach, computing the box, crossed-box and
an additional contact term diagrams (the latter is required for gauge invariance and is
constructed by minimal substitution). 
As the intermediate hadronic states in
the loop diagrams, we include a nucleon and a $\Delta$. The two-photon exchange effects from the intermediate nucleon and $\Delta$ have opposite sings in most kinematical regimes,
being pronounced even at low energies. 
An important theoretical ingredient of the calculation is the $\gamma \Delta \Delta$
vertex. Various forms of this vertex feature prominently in the studies of electromagnetic interactions of the deuteron~\cite{Are04,Are74} and  
three-nucleon bound states~\cite{Del04}, as well as in the recent extraction~\cite{Pas05} of the $\Delta$ magnetic dipole moment.  
In the present calculation we will show that the angular nonlinearity of the two-photon exchange correction is rather sensitive to the value and sign of the 
(dominant) $\gamma \Delta \Delta$ magnetic coupling constant. 
Being formulated in terms of hadronic degrees of freedom, our model is somewhat
complementary to the approach of Ref.~\cite{Pas06} where two-photon exchange effects were calculated using the formalism of generalised parton distributions.

\section{Cross section}\seclab{csec}

The differential cross section for the unpolarized $\Delta$ production process in the
electron-proton collisions can be written as 
\beq
\sigma = \sigma_B (1+\delta_{2 \gamma})\,, 
\eqlab{cs}
\eeq
where $\sigma_B$ is the
one-photon exchange (Born) cross section and $\delta_{2 \gamma}$ is a
two-photon exchange correction. We do not recapitulate in this paper the calculation of the standard radiative corrections such as the vacuum polarization, electron self-energy and vertex loops, as well as the soft-photon bremsstrahlung emission. The derivation of these 
contributions is well-known~\cite{MoT69} and they are taken into account in data 
analyses~\cite{Lia04,Tva05}. In particular, the soft-photon emission 
is necessary to obtain an infrared convergent amplitude; this contribution is
included in our calculation as will be explained in \secref{loop_int}.
As two independent kinematical variables
we choose $Q^2 \equiv -q^2 \ge 0$, the square of the momentum transferred in the collision, and
$\epsilon$, the longitudinal polarization of the virtual photon exchanged in the Born approximation.
In the laboratory frame, $\epsilon$ is related to the electron scattering angle $\theta$ through
$\epsilon^{-1}=1+2 (\tan^2 \theta/2) (\nu^2+Q^2)/Q^2$, 
where $\nu=(M_\Delta^2-M^2+Q^2)/(2 M)$, with $M$ and $M_\Delta$ the nucleon and 
$\Delta$  masses, respectively. 

To leading order in the electromagnetic coupling constant, the two-photon exchange  correction is evaluated from
\beq
\delta_{2 \gamma} = 2 \frac{\mbox{Re} \left(
{\mathcal{M}}_B^\dagger {\mathcal{M}}_{2 \gamma} 
\right)}{\left| {\mathcal{M}}_B \right|^2} \;,
\eqlab{del}
\eeq
where ${\mathcal{M}}_B$ and ${\mathcal{M}}_{2 \gamma}$ denote the Born
and the two-photon exchange scattering amplitudes, respectively. 
The loop diagrams for the two-photon exchange amplitude, as calculated in~\secref{loop_int}, are shown in~\figref{dprod}. These include the contributions of 
nucleon and $\Delta$ intermediate states, plus an additional contact term necessary
to ensure gauge invariance of the full amplitude (see~\secref{gaugeinv}).

\section{Hadronic vertex functions and propagators}\seclab{delver}

The loop diagrams in~\figref{dprod} consist of the propagators of the intermediate particles,
three-point hadronic vertices and a four-point contact term. The construction of the
contact term is explained in~\secref{gaugeinv}; the three-point vertices and
two-point propagators are described in this section.

The $\gamma N N$ vertex is chosen in the standard form, depending only on the
photon four-momentum $q$:
\beq
\Gamma_\mu(q) = 
\gamma_\mu F_1(q^2) + i\frac{\sigma_{\mu \nu} q^\nu}{2 M} F_2(q^2)  \,,
\eqlab{nngam}
\eeq
where $F_{1,2}(q^2)$ denote the Dirac and Pauli form factors,
respectively.~\footnote{We use the conventions and notation of Ref.~\cite{Bjo64}.}

The $\gamma N \Delta$ vertex is taken as~\cite{Kon05}
\beq
V^{\alpha \mu}(p,q) =  \frac{F_{\gamma N \Delta}(q^2)}{2 M_{\Delta}^2} \bigg\{
g_1 \left[\, g^{\alpha \mu} \qslash{q} \vslash{p} - \qslash{q} \gamma^\alpha p^\mu
- \gamma^\alpha \gamma^\mu p \cdot q + q^\alpha \vslash{p} \gamma^\mu  \, \right]  
+ g_2 \left[\, q^\alpha p^\mu - g^{\alpha \mu} p \cdot q\, \right]  \bigg\} \gamma_5\,, 
\eqlab{ndgam}
\eeq
where $q_\mu$ and $p_\alpha$ are the four-momenta of an incoming photon and an outgoing 
$\Delta$, respectively, and the coupling constants 
$g_1$ and $g_2-g_1$ correspond to the magnetic and electric components of the vertex.
(for simplicity, the strongly suppressed Coulomb component is not included in this calculation).
As in the $\gamma N N$ vertex, the $\gamma N \Delta$ form factor $F_{\gamma N \Delta}(q^2)$
is taken as a function of the photon four-momentum squared only.

We choose the $\gamma \Delta \Delta$ vertex in the following form: 
\beq
\Gamma_{\alpha \beta}^\mu(p',p;q)=
F_{\gamma \Delta \Delta}(q^2) \frac{g_\Delta}{M_\Delta^5} p^{\prime 2} p^2 
{\mathcal{P}}^{3/2}_{\alpha \kappa}(p') 
\left( g^{\kappa \mu} q^\eta - g^{\mu \eta} q^\kappa \right)
{\mathcal{P}}^{3/2}_{\eta \beta}(p) \,,
\eqlab{ddgam}
\eeq
where $g_\Delta$ is the coupling constant, and $p^{\prime}_\alpha$, $p_\beta$ and $q_\mu$ are the four-momentum components of the
outgoing $\Delta$, incoming $\Delta$ and incoming photon, respectively, and 
the spin $3/2$ projection operator equals
\beq
{\mathcal{P}}^{3/2}_{\alpha \beta}(p) = g_{\alpha \beta} - \frac{\gamma_\alpha \gamma_\beta}{3}-
\frac{\vslash{p} \gamma_\alpha p_\beta + p_\alpha \gamma_\beta \vslash{p}}{3 p^2} \,.
\eqlab{proj}
\eeq 
Note that the most general form of the $\gamma \Delta \Delta$ vertex contains
several independent structures. However, the coupling constants in the general vertex are hard to extract from data and hence are subject to some uncertainty at present~\cite{Are04,Pas05}. Therefore, to keep the computations tractable, we have 
chosen in~\eqref{ddgam} a form of the $\gamma \Delta \Delta$ vertex
containing only a magnetic dipole component,
which is expected to be dominant~\cite{Are04,Are74}. Other structures are omitted;  
in particular, a component proportional solely to the electric charge of the $\Delta$ is not included in~\eqref{ddgam}. 

Even though the $\gamma N \Delta$ and $\gamma \Delta \Delta$ vertices used in this
calculation are not of the most general form, it is important that they posses the following
transversality properties:
the scalar products of both vertices~\eqref{ndgam} and~\eqref{ddgam} with the four-momenta of the photon or either $\Delta$ vanish.
The photon transversality is necessary for electromagnetic gauge invariance~\cite{Wei95} of the calculation.
The orthogonality to the $\Delta$ four-momentum ensures propagation of only the physical  component of the $\Delta$, thus allowing us to use only the spin $3/2$ part 
$S^\Delta_{\alpha \beta}(p) = (\vslash{p}-M_\Delta)^{-1}
{\mathcal{P}}^{3/2}_{\alpha \beta}(p)$
of the Rarita-Schwinger propagator (see Ref.~\cite{Pas98} for 
a discussion of this property). For the nucleon we take
the standard free fermion propagator $S(p)=(\vslash{p}-M)^{-1}$.

The nonlocality of the hadron-photon interactions in
Eqs.~(\ref{eq:nngam}, \ref{eq:ndgam}) and (\ref{eq:ddgam}) 
is described by the form factors $F_{1,2}$, $F_{\gamma N \Delta}$ and 
$F_{\gamma \Delta \Delta}$. The form factors are necessary to ensure ultraviolet convergence of the loop integrals calculated in~\secref{loop_int}.
The results presented below were obtained using the standard
dipole form for all these form factors:
\beq  
F(q^2)=\frac{\Lambda^4}{(\Lambda^2-q^2)^2},
\eqlab{ff}
\eeq
where $\Lambda$ is the cutoff mass. 
More details on the choice of the form factor will be given in~\secref{res}.

\section{Gauge invariance and the $\gamma \gamma N \Delta$ contact term}\seclab{gaugeinv}

The requirement of electromagnetic gauge invariance can be formulated for the
two-photon exchange graphs in~\figref{dprod} in the following manner. First we ``strip off" the electron line together with the attached photon propagators and thus
consider only the remaining hadron part. For the purposes of this section, 
the four-momentum of the 
incoming proton will be denoted as $p$, that of the 
outgoing $\Delta$ as $p'_\alpha$, and those of the initial and final photons as 
$q_\nu$ and $q'_\mu$, respectively (both photons incoming), so that 
the four-momentum conservation in this notation reads $p'=p+q+q'$. We denote the thus obtained hadronic amplitude as ${\mathcal{M}}_{\alpha \mu \nu}(q',q;p',p)$. Gauge invariance then leads to the orthogonality conditions (see, e.~g.,~\cite{Wei95})
\beq
q^{\prime \mu} {\mathcal{M}}_{\alpha \mu \nu}(q',q;p',p)=
q^{\nu} {\mathcal{M}}_{\alpha \mu \nu}(q',q;p',p)=0 \,.
\eqlab{gi}
\eeq 

First we note that Eqs.~(\ref{eq:gi}) are not satisfied by the sum 
${\mathcal{M}}^{box}_{\alpha \mu \nu}(q',q;p',p) +
{\mathcal{M}}^{xbox}_{\alpha \mu \nu}(q',q;p',p)$, where
\beq
{\mathcal{M}}^{box}_{\alpha \mu \nu }(q',q;p',p)=
V_{\alpha \mu}(q',p') S(p+q) \Gamma_\nu(q) \,, 
\eqlab{box_hadr}
\eeq
\beq
{\mathcal{M}}^{xbox}_{\alpha \mu \nu}(q',q;p',p)=
V_{\alpha \nu}(q,p') S(p+q') \Gamma_\mu(q') \,,
\eqlab{xbox_hadr}
\eeq
are the box and crossed-box hadronic amplitudes with a nucleon intermediate state. 
This can be verified by using the hadron-photon vertices
Eqs.~(\ref{eq:nngam}--\ref{eq:ddgam}), 
the free nucleon propagator $S(p)$ and the
Dirac equation for the incoming nucleon and outgoing $\Delta$.
To obtain a gauge invariant amplitude, an additional $\gamma \gamma N \Delta$ contact term 
${\mathcal{M}}^{ct}_{\alpha \mu \nu}(q',q;p',p)$ 
(depicted by the triangle in~\figref{dprod})
has to be added to the box and crossed-box
diagrams. 

Such a term can be constructed by the standard procedure of minimal 
substitution (the technical details can be found for example in Ref.~\cite{Kon00}). 
An inherent property of the minimal substitution is that it yields unambiguously only the contact term longitudinal to the photons' four-momenta (which is sufficient for current conservation).
Following the procedure of Ref.~\cite{Kon00}
we obtain the contact term
\beq
{\mathcal{M}}^{ct}_{\alpha \mu \nu}(q',q;p',p)=
-\frac{2 p_\nu + q_\nu}{(p+q)^2-p^2} F_1(q^2) V_{\alpha \mu}(p',q') -
\frac{2 p_\mu + q'_\mu}{(p+q')^2-p^2} F_1(q^{\prime 2}) V_{\alpha \nu}(p',q) \,.
\eqlab{ct_hadr}
\eeq
It can be verified now that the sum of the nucleon box, crossed-box and contact term amplitudes
\beq
{\mathcal{M}}_{\alpha \mu \nu}(q',q;p',p)={\mathcal{M}}^{box}_{\alpha \mu \nu}(q',q;p',p) +
{\mathcal{M}}^{xbox}_{\alpha \mu \nu}(q',q;p',p)+
{\mathcal{M}}^{ct}_{\alpha \mu \nu }(q',q;p',p) \,,
\eqlab{ampl}
\eeq
obeys~\eqref{gi}, i.~e.~the upper row of~\figref{dprod} is gauge invariant.

Finally, using the vertices and propagators given in~\secref{delver}, it is easy to see that the sum of the box and crossed-box 
amplitudes with an intermediate $\Delta$ (the lower part of~\figref{dprod}) 
is gauge invariant by itself.
Therefore the sum of the loop diagrams depicted in~\figref{dprod} constitute a 
gauge invariant two-photon exchange amplitude.

\section{Loop integrals}\seclab{loop_int}
 
The loop integrals corresponding to the box and crossed-box 
diagrams with an intermediate nucleon in~\figref{dprod} can be written as
\beq 
{\mathcal{M}}_N^{\gamma \gamma} = 
e^4\!\int \! \frac{d^4 k}{(2 \pi)^4}
\frac{N^N_{box}(k)}{D^N_{box}(k)} +
e^4\!\int \! \frac{d^4 k}{(2 \pi)^4} 
\frac{N^N_{xbox}(k)}{D^N_{xbox}(k)} \,, 
\eqlab{integr_n}
\eeq
with the numerators and denominators given by
\bea
N^N_{box}(k) &=& 
\frac{2}{3} \oln{U}_\alpha(p_4) V^{\alpha \mu}(p_4,q-k) 
\left[ \vslash{p}_2+\vslash{k}+M \right] 
\Gamma^{\nu}(k) U(p_2) \nn \\
& \times &
\oln{u}(p_3) \gamma_\mu \left[ \vslash{p}_1 -\vslash{k} \right] \gamma_\nu u(p_1)\,, 
\eqlab{num_n_box} \\
N^N_{xbox}(k) &=& 
\frac{2}{3}\oln{U}_\alpha(p_4) V^{\alpha \mu}(p_4,q-k)
\left[ \vslash{p}_2+\vslash{k}+M \right] 
\Gamma^{\nu}(k) U(p_2) \nn \\
& \times &
\oln{u}(p_3) \gamma_\nu \left[ \vslash{p}_3 +\vslash{k} \right] \gamma_\mu u(p_1)\,, 
\eqlab{num_n_x-box} 
\eea
\bea
D^N_{box}(k) & = & k^2 (k-q)^2 \left[ (p_1-k)^2-m^2 \right] \left[ (p_2+k)^2-M^2 \right] , 
\eqlab{den_n_box} \\
D^N_{xbox}(k) & = & \left. D^N_{box}(k) \right|_{p_1-k \rightarrow p_3+k}\,.
\eqlab{den_n_x-box} 
\eea
Here $U_\alpha$ is the $\Delta$ vector-spinor, and $U$ and $u$ are the proton and electron spinor wave functions, respectively (the electron mass $m$ has been neglected in the numerators).

The loops with an intermediate $\Delta$ can be written
\beq 
{\mathcal{M}}_\Delta^{\gamma \gamma} = 
-e^4\!\int \! \frac{d^4 k}{(2 \pi)^4}
\frac{N^\Delta_{box}(k)}{D^\Delta_{box}(k)} -
e^4\!\int \! \frac{d^4 k}{(2 \pi)^4} \frac{N^\Delta_{xbox}(k)}{D^\Delta_{xbox}(k)} \,, 
\eqlab{integr_d}
\eeq
where 
\bea
N^\Delta_{box}(k) &=& 
\frac{2}{3} \oln{U}_\alpha(p_4) \Gamma_{\alpha \kappa}^{\mu}(p_4,p_2+k,q-k) 
\left[ \vslash{p}_2+\vslash{k}+M_\Delta \right] {\mathcal{P}}^{3/2}_{\kappa \beta}(p_2+k)
V^{\beta \nu}(p_2+k,k) U(p_2) \nn \\
& \times &
\oln{u}(p_3) \gamma_\mu \left[ \vslash{p}_1 -\vslash{k} \right] \gamma_\nu u(p_1)\,, 
\eqlab{num_d_box} \\
N^\Delta_{xbox}(k) &=& 
\frac{2}{3} \oln{U}_\alpha (p_4) \Gamma_{\alpha \kappa}^{\mu}(p_4,p_2+k,q-k) 
\left[ \vslash{p}_2+\vslash{k}+M_\Delta \right] {\mathcal{P}}^{3/2}_{\kappa \beta}(p_2+k)
V^{\beta \nu}(p_2+k,k) U(p_2) \nn \\
& \times &
\oln{u}(p_3) \gamma_\nu \left[ \vslash{p}_3 +\vslash{k} \right] \gamma_\mu u(p_1)\,, 
\eqlab{num_d_x-box} 
\eea
\bea
D^\Delta_{box}(k) & = & k^2 (k-q)^2 \left[ (p_1-k)^2-m^2 \right] 
\left[ (p_2+k)^2-M_\Delta^2 \right] , 
\eqlab{den_d_box} \\
D^\Delta_{xbox}(k) & = & \left. D^\Delta_{box}(k) \right|_{p_1-k \rightarrow p_3+k}\,. 
\eqlab{den_d_x-box} 
\eea

The loop with the $\gamma \gamma N \Delta$ contact term reads
\beq 
{\mathcal{M}}_{ct}^{\gamma \gamma} = 
-e^4\!\int \! \frac{d^4 k}{(2 \pi)^4}
\frac{N_{ct}(k)}{D_{ct}(k)} \,, 
\eqlab{integr_ct}
\eeq
where
\bea
N_{ct}(k) &=& 
\frac{2}{3} \oln{U}_\alpha(p_4) {\mathcal{M}}_{ct}^{\alpha \mu \nu}(q-k,k;p_4,p_2) U(p_2) \nn 
\\
& \times &
\oln{u}(p_3) \gamma_\mu \left[ \vslash{p}_1 -\vslash{k} \right] \gamma_\nu u(p_1)\,, 
\eqlab{num_ct}  
\eea
\bea
D_{ct}(k) & = & k^2 (k-q)^2 \left[ (p_1-k)^2-m^2 \right]\,.
\eqlab{den_ct} 
\eea

The full amplitude, comprising 
the sum of the above loop integrals, is both infrared and ultraviolet convergent.
The ultraviolet convergence is provided through the presence of regularising form factors in
the integrands of Eqs.~(\ref{eq:integr_n},\ref{eq:integr_d}) and (\ref{eq:integr_ct}).
The infrared divergence from the box and crossed-box diagrams with an intermediate nucleon is obtained by putting $k \rightarrow 0$ 
in Eqs.~(\ref{eq:num_n_box}) and (\ref{eq:num_n_x-box}) (these integrals converge at 
$k \rightarrow q$). This divergence is cancelled in the standard way against the corresponding divergence of the
soft-photon emission contributions~\cite{MoT69,Blu03}. In our approach, the latter enter
as part of the loop diagram with the contact term, obtained from the divergences
of~\eqref{integr_ct}
at $k \rightarrow 0$ (to cancel the divergence of the box integral) and 
$k \rightarrow q$ (to cancel the divergence of the crossed-box integral).
We incorporate the soft-photon divergence into the contact term merely for convenience, so that in addition to restoring gauge invariance, the contact term also ensures the infrared convergence. As an additional check, we
confirmed numerically that the full amplitude is indeed convergent.

We calculated the loop integrals Eqs.~(\ref{eq:integr_n},\ref{eq:integr_d},\ref{eq:integr_ct}) using techniques similar to those described in~\cite{Blu03,Kon05}.
The result of the integration is obtained analytically
as a sum of several Passarino-Veltman dilogarithms~\cite{tHo79}, each 
multiplied by a rational function of kinematic invariants. 
Although simple in structure, the explicit expressions are
quite lengthy and therefore would not be useful within the limits of this paper. 
We checked that the calculation obeys the constraint of crossing symmetry, i.~e.~the amplitude is invariant under an interchange of the Mandelstam variables, as explained 
in~\cite{Kon05}. 
The conditions of crossing symmetry and gauge invariance are
important because they put strong model independent constraints on the behaviour of the
amplitudes.

\begin{figure}[!htb]
\centerline{{\epsfxsize 10.0cm \epsffile[15 250 570 520]{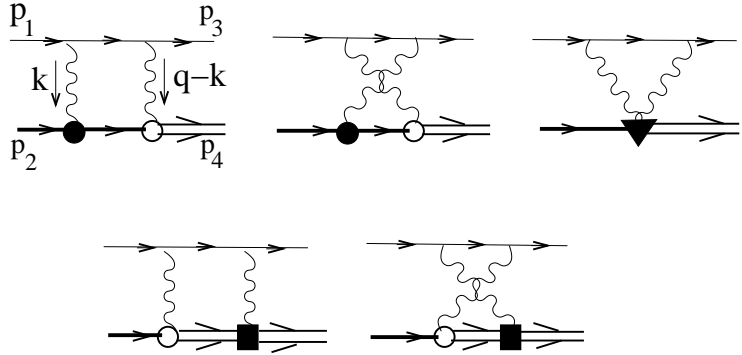}}}
\caption[f1]{Two-photon exchange graphs for the 
$e p \rightarrow e \Delta$ reaction. The thin and thick straight lines denote electrons and
protons, respectively; the double lines are $\Delta$s, the wavy lines photons. 
In the box and crossed-box diagrams, the black squares, 
black circles and white circles denote, respectively, the $\gamma \Delta \Delta$, 
$\gamma N N$ and $\gamma N \Delta$ vertices used in the calculation. 
The loop diagram with the $\gamma \gamma N \Delta$ contact term (denoted by the 
black triangle) ensures gauge invariance of the calculation.
\figlab{dprod}}
\end{figure}

\section{Results}\seclab{res}

We did several calculations of the two-photon exchange correction using
various functional forms of the regularizing form factor in the hadronic vertices,
including dipole and monopole functions with several values of the cutoff. 
We found that the
essential features of the results are largely independent of the 
details of the form factor. This is because the
form factor is partially cancelled when
$\delta_{2 \gamma}$ is expressed as the ratio~\eqref{del}. 

As a representative example, we show in~\figref{dpl}
the two-photon exchange correction 
which was obtained using the dipole form factor~\eqref{ff}
with the cutoff $\Lambda=0.84$ GeV. This choice is known to be compatible with the nucleon mean-square radius. 
Changing the form factor (for example, using a monopole cutoff of a similar size)
has no significant effect on $\delta_{2 \gamma}$ over most of the kinematic range in
$\epsilon$, with only a mild influence for $\epsilon \approx 1$. 

As mentioned above, the $\gamma \Delta \Delta$ vertex is
an important ingredient of this calculation. Recent experimental analyses~\cite{Tva05} of Rosenbluth separations
focused on the possibility to identify two-photon exchange effects through a nonlinearity of the cross section as a function of $\epsilon$. The nonlinear behaviour of
the two-photon exchange corrections was also addressed in theoretical studies.
Thus, general properties of two-photon exchange corrections in elastic
electron-proton scattering were derived in Ref.~\cite{Rek04}. Nonlinearities 
induced by two-photon exchanges were searched for in the data for 
electron-deuteron~\cite{Rek99} and electron-proton~\cite{Tom05} scattering.

In the present calculation of $\Delta$ production in
{\em inelastic} electron-proton collisions
we find a pronounced dependence of the two-photon correction on the value of the
magnetic $\gamma \Delta \Delta$ coupling constant $g_\Delta$. 
As shown in~\figref{dpl}, the sign of $g_\Delta$ is related to the qualitatively distinct nonlinearities of $\delta_{2 \gamma}$ in $\epsilon$. An essentially linear behaviour of the 
two-photon correction is consistent with $g_\Delta = 0$, corresponding to 
the presence of only a nucleon intermediate state and no intermediate $\Delta$. 
The current analysis~\cite{Tva05} of experimental Rosenbluth separations in electron-proton 
collisions indicated only a very small nonlinearity, although additional data should allow one to reach a more definite conclusion about the nonlinearities. 

\begin{figure}[!htb]
\centerline{{\epsfxsize 12.0cm \epsffile[15 330 570 740]{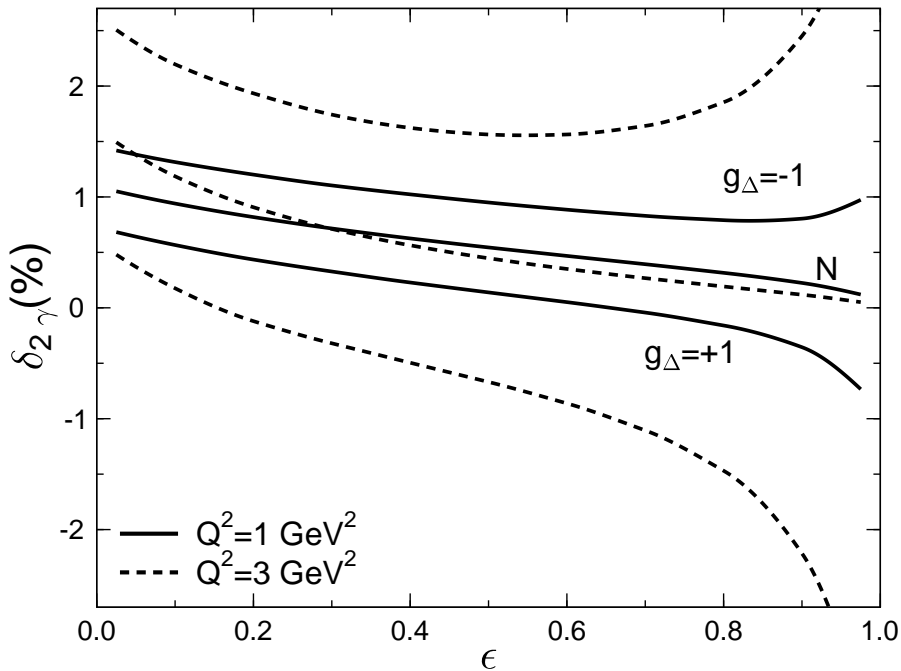}}}
\caption[f2]{Two-photon exchange correction to the unpolarized cross section for
$\Delta$ production in electron-proton collisions, calculated for 
$Q^2=1$ GeV$^2$ (solid lines) and $Q^2=3$ GeV$^2$ (dashed lines). 
The upper, middle and lower pairs of lines are labeled by
the values of the $\gamma \Delta \Delta$ coupling constant $g_{\Delta}$ 
(``N" corresponding to $g_\Delta=0$, i.~e.~to the absence of an intermediate $\Delta$ state). 
\figlab{dpl}}
\end{figure}

\section{Concluding remarks} \seclab{concl}

An accurate description of electron-proton scattering in the resonance region
is essential for the understanding of the structure and interactions of hadrons. 
Using Born approximation to analyze $\Delta$ production in electron-proton scattering, one 
can obtain important information of the $\gamma N \Delta$ transition current.
However, such an analysis should also include two-photon exchange 
higher-order corrections. The two-photon exchange corrections calculated in this paper are generally much smaller than the Born contribution, 
as should be expected from perturbation theory. Nevertheless, our results show that 
the two-photon
exchange corrections can be quite important for a precise analysis of the
electron-proton scattering in the resonance region. 

For further progress in the evaluation of higher-order effects in
electron-nucleon collisions, a more detailed knowledge 
of the $\gamma \Delta \Delta$ vertex is needed.  
We have shown that 
the nonlinearity of the $\epsilon$ dependence of the two-photon correction
can be related to the sign of the magnetic $\gamma \Delta \Delta$ coupling constant. 
In particular, a
linear dependence would suggest that the contribution of an intermediate $\Delta$ state to the two-photon exchange correction
is negligible in comparison with that of the nucleon intermediate state.
The question whether
the link between the magnetic $\gamma \Delta \Delta$ coupling and the $\epsilon$ nonlinearity might be affected by using more general
forms of the $\gamma \Delta \Delta$ and $\gamma N \Delta$ vertices falls outside the scope of this paper, but should be addressed in more detailed studies in the future. 
Another interesting problem to be addressed in the future is the role 
of higher mass resonances in the intermediate states, 
in addition to the essential contribution of the nucleon and 
$\Delta$ considered in this paper.

Finally, we point out that the present calculation is one of the few existing theoretical 
studies of the two-photon exchange effects in the
$\Delta$ production reaction; therefore,   
additional investigations would certainly be useful and interesting.
         
\begin{acknowledgments}
We thank Wally Melnitchouk and Vladis Tvaskis for useful discussions.
\end{acknowledgments}



\begin{thebibliography}{99}
\bibitem{Dre59} S.~D.~Drell and S.~Fubini, \PR{113}{1959}{741};                
                Y.-S.~Tsai, \PR{122}{1961}{1898}; 
                G.~K.~Greenhut, \PR{184}{1969}{1860}; 
                V.~N.~Boitsov, L.~A.~Kondratyuk, and V.~B.~Kopeliovich,
                Sov.~J.~Phys.~{\bf16}, 287 (1973);
                J.~Guinon and L.~Stodolsky, \PRL{30}{1973}{345}. 
\bibitem{MoT69} L.~Mo and Y.-S.~Tsai, Rev.~Mod.~Phys.~{\bf41}, 205 (1969);
                L.~C.~Maximon and J.~A.~Tjon, \PRC{62}{2000}{054320}.
\bibitem{Blu03} P.~G.~Blunden, W.~Melnitchouk, and J.~A.~Tjon, \PRL{91}{2003}{142304};
                P.~G.~Blunden, W.~Melnitchouk, and J.~A.~Tjon, \PRC{72}{2005}{034612}.
\bibitem{Gui03} P.~A.~M.~Guichon and M.~Vanderhaeghen, \PRL{91}{2003}{142303}.
\bibitem{Che04} Y.~C.~Chen, A.~Afanasev, S.~J.~Brodsky, C.~E.~Carlson 
                and M.~Vanderhaeghen,  \PRL{93}{2004}{122301}.
\bibitem{Kon05} S.~Kondratyuk, P.~G.~Blunden, W.~Melnitchouk, and J.~A.~Tjon,
                \PRL{95}{2005}{172503}.
\bibitem{Lia04} Y.~Liang et al., nucl-ex/0410027.
\bibitem{Tva05} V.~Tvaskis et al., \PRC{73}{2006}{025206}.
\bibitem{Pas06} V.~Pascalutsa, C.~E.~Carlson and M.~Vanderhaeghen, \PRL{96}{2006}{012301}.
\bibitem{Are04} H.~Arenh{\"o}vel, W.~Leidemann, E.~L.~Tomusiak, nucl-th/0407053,
\bibitem{Are74} H.~Arenh{\"o}vel and H.~G.~Miller, \ZP{266}{1974}{13};
                H.~J.~Weber and H.~Arenh{\"o}vel, Phys.~Rep.~{\bf36}, 277 (1978).
\bibitem{Del04} A.~Deltuva, L.~P.~Yuan, J.~Adam Jr., A.~C.~Fonseca, and P.~U.~Sauer,
                \PRC{69}{2004}{034004}.
\bibitem{Pas05} V.~Pascalutsa and M.~Vanderhaeghen, \PRL{94}{2005}{102003}.  
\bibitem{Bjo64} J.~D.~Bjorken and S.~D.~Drell,
                {\it Relativistic Quantum Mechanics} (McGraw-Hill, 1964).
\bibitem{Wei95} S.~Weinberg, The Quantum Theory of Fields, vol.~1 
                (Cambridge University Press, 1995).
\bibitem{Pas98} V.~Pascalutsa, \PRD{58}{1998}{096002};
                V.~Pascalutsa and R.~Timmermans, \PRC{60}{1999}{042201}. 
\bibitem{Kon00} S.~Kondratyuk and O.~Scholten, \NPA{677}{2000}{396}.
\bibitem{tHo79} G.~'t Hooft and M.~Veltman, \NPB{153}{1979}{365}; 
                G.~Passarino and M.~Veltman, \NPB{160}{1979}{151}.
\bibitem{Rek04} M.~P.~Rekalo and E.~Tomasi-Gustafsson, Eur.~Phys.~J.~A~{\bf22}, 331 (2004).
\bibitem{Rek99} M.~P.~Rekalo, E.~Tomasi-Gustafsson, and D.~Prout, \PRC{60}{1999}{04202}.
\bibitem{Tom05} E.~Tomasi-Gustafsson and G.~I.~Gakh, \PRC{72}{2005}{015209}.
\end{thebibliography}
\end{document}